\documentclass[useAMS,usenatbib]{mn2e}
\usepackage{psfig, epsf, epsfig}
\title[On the origin of double
main-sequence turn-offs]{On the origin of double 
main-sequence turn-offs in star clusters of the Magellanic Clouds}
\author[K. Bekki
and A. D. Mackey ]{Kenji Bekki${}^1$\thanks{E-mail:
bekki@phys.unsw.edu.au} and
 A. D. Mackey${}^2$ \\
       ${}^1$School of Physics, University of New South Wales,
              Sydney 2052, NSW, Australia\\
      ${}^2$Institute for Astronomy, University of Edinburgh, Royal
Observatory, Blackford Hill, Edinburgh EH9 3HJ}

\begin{document}

\date{Accepted --, Received --; in original form --}

\pagerange{\pageref{firstpage}--\pageref{lastpage}} \pubyear{2008}

\maketitle

\label{firstpage}

\begin{abstract}

Recent observational studies of intermediate-age star clusters (SCs)
in the Large Magellanic Cloud (LMC)
have reported that a significant number of these objects show double 
main-sequence turn-offs (DMSTOs) in their color-magnitude diagrams 
(CMDs). One plausible explanation for the origin
of these DMSTOs is that the SCs are composed
of two different stellar populations with age differences of $\sim 300$ Myr.
Based on analytical methods and numerical simulations,
we explore a new scenario in which 
SCs interact and merge with star-forming giant molecular clouds (GMCs)
to form new composite SCs with two distinct component populations.
In this new scenario, the possible age differences between the
two different stellar populations responsible for the DMSTOs
are due largely to secondary star formation within GMCs interacting 
and merging with already-existing SCs in the LMC disk.
The total gas masses being converted into new stars
(i.e., the second generation of stars)
during GMC-SC interaction and 
merging can be comparable to or larger than the masses
of the original SCs (i.e, the first generation of stars) in this scenario.
Our simulations show that the spatial distributions of new stars 
in composite SCs formed from GMC-SC merging are more compact than 
those of stars initially in the SCs. 
We discuss both advantages and disadvantages of the new scenario in
explaining
fundamental properties of SCs with DMSTOs in the LMC and in the Small
Magellanic Cloud (SMC). 
We also discuss the merits of various alternative scenarios for the 
origin of the DMSTOs.

\end{abstract}


\section{Introduction}

Recent photometric studies of intermediate-age SCs in the LMC
have discovered objects exhibiting extremely unusual main-sequence 
turn-offs (MSTOs) in their CMDs. The first indications that
such clusters might exist date back several years to observations 
obtained with terrestrial facilities -- Bertelli et al. (2003)
demonstrated that the LMC cluster NGC 2173 apparently possesses 
an unusually large spread in colour about its MSTO, while Baume 
et al. (2007) obtained a similar result for NGC 2154.

More recent studies based on deep precision photometry from the 
{\it Hubble Space Telescope} ({\it HST}) Advanced Camera for 
Surveys (ACS) have revealed the truly peculiar nature of many
intermediate-age LMC clusters. Most striking are the relatively 
massive clusters NGC 1846, 1806, and 1783 which   
possess double MSTOs (hereafter referred to as DMSTOs
for convenience) on their CMDs (Mackey \& Broby Nielsen 2007,
MBN07; Mackey et al. 2008; Milone et al. 2008; Goudfrooij et al.
2008). Despite this, the remaining CMD sequences for these
clusters (i.e., their red-giant branches, subgiant branches,
main sequences, and red clumps) are all extremely narrow and
well-defined. This suggests that none of the clusters possesses
a significant line-of-sight depth or internal dispersion in
$[$Fe$/$H$]$, or suffers from significant differential extinction.
The simplest explanation is that the observed DMSTOs in each of
these clusters represent two distinct stellar populations with
differences in age of $\sim 200 - 300$ Myr. It should be noted,
however, that apart from the study of Mucciarelli et al. (2008)
who found no significant star-to-star dispersion in $[\alpha/$Fe$]$
for $6$ stars in NGC 1783, no constraints on the possibility of 
internal variations in chemical abundances other than $[$Fe$/$H$]$
presently exist for these SCs.

Milone et al. (2008) studied a large sample of $16$ intermediate-age
LMC clusters, and found that $11$ of these possess CMDs exhibiting
MSTOs which are not consistent with being simple, single stellar
populations. Of these $11$, four clearly show DMSTOs (the three
described above plus NGC 1751) while the remainder possess more
sparsely populated CMDs that make the precise morphologies of their
peculiar MSTOs difficult to ascertain. All are consistent with
being DMSTOs, or alternatively with possibly being more smoothly
distributed intrinsic broadenings. Nonetheless, all $11$ clusters again
exhibit very narrow sequences across the remainder of their CMDs,
suggesting that their peculiar MSTOs are due to internal age 
dispersions of $\sim 200-300$ Myr. 
Milone et
al. (2008) measured  the ratio between  the  upper and the  lower MSTO
population in   NGC1846, NGC1806 and   in  the less-populated NGC1751
and found that the inferred young 
populations may comprise up to $\sim 70$ per cent of the stars in 
the central regions of these SCs.

In addition to the above objects, several younger LMC clusters
have been identified as possibly possessing irregular CMDs
(e.g., Santiago et al. 2002, Gouliermis et al. 2006).
It is further worth noting that one intermediate-age SMC cluster, 
NGC 419, also likely possesses a DMSTO (Glatt et al. 2008).

The results described above suggest that, contrary to expectation,
DMSTOs may be a common feature among intermediate-age SCs of
the Magellanic Clouds (MCs).
If the origin of such DMSTOs is indeed closely associated with
an age difference between a first and second generation
of stars in these SCs, 
the above observations raise the following three key questions: 
(i) why the second generation of stars in a SC with a DMSTO
can be formed about 300 Myr after the formation of the first
generation of stars, (ii) why the total mass of the second 
generation can be comparable or even larger than that of the first,
and (iii) how the two different populations can now be in the same
SC. 
Although the origin of the observed multiple stellar populations 
in the Galactic globular clusters (GCs) has been observationally 
and theoretically discussed in terms of GC formation scenarios 
(e.g., Piotto 2008),
the above fundamental questions so far have not been discussed
by theoretical models of SC formation in the LMC.

\begin{table*}
\centering
\begin{minipage}{175mm}
\caption{Model parameters for numerical simulations of GMC-SC mergers}
\begin{tabular}{ccccccc}
{Model no 
\footnote{For models M3 - M32, the ranges of model parameters
investigated in the present study are shown in parentheses: the left and
the right numbers are the minimum and maximum values.}}
& {$M_{\rm s}$
\footnote{The total mass of stars in 
a SC  in units of $10^{4}$  ${\rm M_{\odot}}$. The total gas mass
in a GMC is $5\times 10^4 {\rm M}_{\odot}$ for all models. }}
& {$R_{\rm s}$
\footnote{The size of a SC in units of pc. The size of a GMC is set to be
11.9 pc for all models in the present study.}}
& {$y_{\rm g}$ 
\footnote{Initial $y$-position of a GMC with respect to the center
of a SC. This corresponds to the impact parameter of the GMC-SC collision.}}
& {$|u_{\rm g}|$
\footnote{The absolute magnitude
of the initial velocity of a GMC along the $x$-axis in units of km s$^{-1}$,
with respect to
the center of a SC. For all models, $u_{\rm g} \le 0$.}}
& {Star formation
\footnote{``YES'' (``NO'') means that the star formation
model is (is not) included in the simulation. }}
& Comments \\
M1 &  5.0  &  10  & 6.0  & 2.2 &  NO & the standard model \\
M2 &  5.0  &  10  & 6.0  & 2.2 &  YES &  the star formation model  \\
M3-32 &  (1.0, 20.0)  &  (5, 25)  & (0, 23.8)   & (0, 12.8)  & NO 
\& YES  &  \\
\end{tabular}
\end{minipage}
\end{table*}

The purpose of this paper is to explore a new scenario in terms
of the above three fundamental questions on the origin
of the observed DMSTOs.
In the new scenario,
an already formed SC (i.e., the first generation of stars)
can merge or interact with a GMC 
so that a second generation of stars is formed in the GMC.
The second generation can then merge and mix dynamically
with the SC to form a new SC with two stellar populations
of different ages. Thus, the origin of the DMSTOs in SCs of the LMC
results from merging between SCs and GMCs in the new
``GMC-SC-merger''  scenario.
The typical age difference ($t_{\rm gap}$)
of the two populations in SCs required for
explaining the DMSTOs corresponds to 
the typical time scale of merging between
SCs and GMCs of the LMC in the new scenario.
We mainly discuss (i) whether star formation is likely to be enhanced 
during GMC-SC merging, (ii) the mass fraction of the second 
generation of stars relative to the first in a SC and its dependence
on parameters of GMC-SC merging, and (iii) time scales of
GMC-SC merging in the LMC.

The plan of the paper is as follows: In the next section,
we describe our numerical models for GMC-SC merging in the LMC. 
In \S 3, we
present numerical results
mainly concerning the final distributions of gas and new stars in 
the remnants of GMC-SC mergers.
In \S 4, we discuss the origin of the DMSTOs of SCs 
in a more general way.
In this section, we also discuss the advantages and disadvantages
of two alternative scenarios in explaining the physical properties
of SCs with DMSTOs.
We summarize our conclusions in \S 5.

\section{The model}

The most important question in the present study is
whether the masses of the second generations of stars 
(i.e., new stars)
in the remnants of GMC-SC mergers can be as large as those of the first
(i.e., stars initially in SCs).
For the masses of the second generations to be comparable
to those of the first,
a large amount of gas initially in GMCs needs to be accumulated
within merging SCs for star formation.
Although we do investigate ``star formation processes''
within GMCs, the adopted model for star formation is rather 
idealized and phenomenological, and does not allow us to investigate 
{\it subpc-scale} real star formation processes (e.g., the formation 
of individual stars within the cores of small molecular clouds). 
We therefore consider that here it is more important for us 
to (i) investigate how much gas in GMCs can be captured
in SCs during GMC-SC merging for various different model parameters
for the merging
and (ii) suggest that the merger remnants (i.e., new SCs)
with large fractions ($\sim 0.3-0.5$) of gas can finally become SCs with two
different populations with comparable masses.

We adopt the same numerical code as adopted in
our previous numerical simulations on the hydrodynamical evolution
of interstellar gas in interacting and merging
galaxies (Bekki et al. 2002) and the formation of star clusters 
from GMCs in galaxies (Bekki \& Couch 2003). 
Although we can  
investigate gas dynamics in merging GMCs, 
the adopted TREESPH code
does not allow us to precisely describe long-term stellar  
dynamics such as two-body relaxation processes in SCs.
In future work we will discuss the long-term 
evolution of the remnants of
GMC-SC mergers by using appropriate numerical codes such as
{\tt NBODY4} adopted in our previous simulations of 
SC formation (Hurley \& Bekki 2008, HB08). 
In order to understand more clearly the hydrodynamical evolution
of GMC-SC mergers,
we do not include the external tidal field of the MCs
in the present study.

A SC in a GMC-SC merger
is represented by a Plummer model with mass and size
represented by $M_{\rm s}$ and $R_{\rm s}$,  
respectively.
The scale length of the Plummer model
is fixed at $0.2R_{\rm s}$ for all our simulations.
A GMC is assumed to be a Bonner-Ebert isothermal sphere 
(Ebert 1955; Bonner 1956) with a mass of $M_{\rm g}$,
a size of $R_{\rm g}$,  and a temperature of $T_{\rm g}$.
The radial distribution of a Bonner-Ebert isothermal sphere
is determined by
solving the differential equations 
for a given $T_{\rm g}$
of the sphere.
Guided by the observed relation between
mass densities and sizes of GMCs discovered
by Larson's (1981) relation
and the observed typical mass and size of GMCs in 
the Galaxy (e.g., Solomon et al. 1979),
we use the following 
$R_{\rm g}-M_{\rm g}$ relation;
\begin{equation}
R_{\rm g}  
 =40 \times  (\frac{M_{\rm g}}{5 \times 10^5  {\rm M}_{\odot} })^{0.53}
{\rm pc}
\end{equation}
We investigate models with $M_{\rm g}=5 \times 10^4 {\rm
M}_{\odot}$, $R_{\rm g}=11.9$ pc, and $T_{\rm g}$=2.6 K
in the present study. Since the isothermal Bonner-Evert
sphere is adopted, one of key parameters determining
the gas dynamics of GMC-SC merging is the mass-ratio 
($s_{\rm g}=M_{\rm g}/M_{\rm s}$)
of a GMC to a SC in a given model: if we derive parameter
dependences of numerical results on $s_{\rm g}$
for the above $M_{\rm g}$ , these dependences
can be applied for other $M_{\rm g}$.

The initial locations of the SC and the GMC
in the GMC-SC merger
are set to be 
($x_{\rm s}$, $y_{\rm s}$, $z_{\rm s}$) and
($x_{\rm g}$, $y_{\rm g}$, $z_{\rm g}$), respectively.
For convenience, ($x_{\rm s}$, $y_{\rm s}$, $z_{\rm s}$)
=(0,0,0) so that the initial position of the GMC with respect
to the SC 
can be more clearly understood.
The initial velocities of the SC and the GMC
in the $x$-, $y$-, and $z$-directions
are set to be 
($u_{\rm s}$, $v_{\rm s}$, $w_{\rm s}$)
and ($u_{\rm g}$, $v_{\rm g}$, $w_{\rm g}$), respectively,
and ($u_{\rm s}$, $v_{\rm s}$, $w_{\rm s}$)
= (0,0,0) for simplicity.
The orbital plane of the GMC-SC merger is coincident with
the $x$-$y$ plane for all models in the present study
(i.e., $z_{\rm s}=z_{\rm g}=w_{\rm g}=0$).
We adopt $x_{\rm g}=2R_{\rm g}$  for all models
so that GMC and SCs are not initially in direct contact. 
The initial direction of the velocity vector of the GMC
is only a parameter for the orbit of the GMC
owing to the spherically symmetric
distributions of the model GMCs and SCs.
We therefore assume that the initial direction
is parallel to the $x$-axis toward the negative
$x$ (i.e., $u_{\rm g}<0$ and $v_{\rm g}=0$).
Thus $y_{\rm g}$ and $u_{\rm g}$ are free parameters
that determine the orbit of the GMC with respect to the SC.

To summarize, the key model parameters in the present 
simulations are $M_{\rm s}$ (or $s_{\rm g}$), $R_{\rm s}$,
$y_{\rm g}$ (i.e., the impact parameter), and $u_{\rm g}$.

For all models the total number of particles ($N$) is 
40$\,$000, half of which is for stars, and the initial 
gravitational softening length is 0.24 pc.
We choose these particle numbers in order to facilitate
investigation of the long-term evolution of merger remnants 
in the external LMC potential
in our future numerical studies using {\tt NBODY4}, as adopted
in our previous simulations (HB08):
models of SCs using these codes can be undertaken
within a reasonable time scale for $N\sim 10^4-10^5$
without excessive numerical costs (e.g., D'Ercole et al. 2008; HB08).

\begin{figure}
\psfig{file=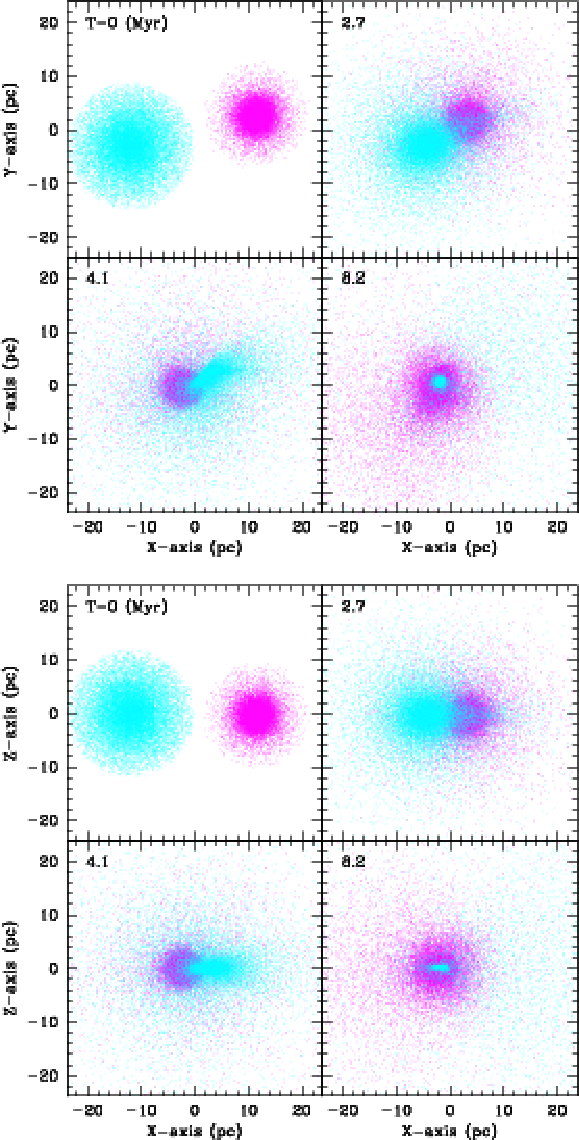,width=8.0cm}
\caption{
Time evolution of stellar (magenta) and gaseous (cyan)
distributions projected onto the $x$-$y$ plane (upper four panels)
and the $x$-$z$ plane (lower four panels) for
the standard model M1. The time $T$ shown in the upper
left corner of each panel is given in units of Myr.
Note that a gaseous disk can be seen in the inner region
of the SC, and the center of the disk seems to deviate from
the center of the SC. 
}
\label{Figure. 1}
\end{figure}

Owing to the limited size and mass resolution
of the present simulations,
subpc-scale physical processes of star formation
cannot be investigated in the present study.
We do, however, try to investigate whether the gas accumulated
in the central regions of SCs after GMC-SC merging
can be sufficient to form new stars by adopting a
simple prescription for star formation.
In the models with ``star formation'',
a gas particle is converted into a collisionless new stellar
one if the gas particle meets the following conditions:
(i) the dynamical time scale of the SPH gas particle
is shorter than the sound crossing time , and (ii) the gas is converging
(i.e., $\nabla {\bf v} <0$, where ${\bf v}$ is the velocity vector of the
gas particle).
These two conditions mimic the Jeans gravitational instability
for gaseous collapse. 
The adopted model for star formation
is rather idealized so that ``new stars'' need to be interpreted
as possible formation sites of stars in GMCs.
In the rest of this paper, the stars initially in SCs are just referred 
to as ``stars'' to distinguish them from new stars formed from gas.

Although we investigate many models spanning different combinations of
the four key parameters, we mainly present the results of the 
``standard model'' which shows the typical behavior of gas accumulation 
in SCs merging with GMCs. 
We also briefly describe the results
of the ``star formation model'' in which the simple treatment of star 
formation is included and which as parameter values exactly the same as 
those adopted in the standard model.  

The parameter values of the standard model and the ranges of
parameters investigated for an additional 31 models
are summarized in Table 1. 
The standard model and the star formation
model are represented as M1  and M2, respectively.
Parameter values in models other than M1
are not described in the table in order to avoid  
consuming an unnecessarily large space for the description.
In the present study we mainly show the dependences
of the gas mass fractions within a radius of 10 pc ($f_{\rm g}$)  
in merger remnants
on $y_{\rm g}$, $u_{\rm g}$, $M_{\rm s}$, and $R_{\rm s}$.

We note that SC-GMC merging does not occur in 
some of our models with large $y_{\rm g}$ and $|u_{\rm g}|$;
however, gas-transfer between SCs and GMCs is still possible
in these cases.
We also describe the results of these models briefly 
in the present paper.
It should be stressed that the present study is the first
step toward understanding the
possibly complicated interactions between SCs and GMCs
in the MCs: the present models are rather idealized in some respects,
in order to grasp the essential ingredients of GMC-SC merging processes.
We plan to investigate a more fully-consistent 
and sophisticated model for GMC-SC merging in our future studies.

\begin{figure}
\psfig{file=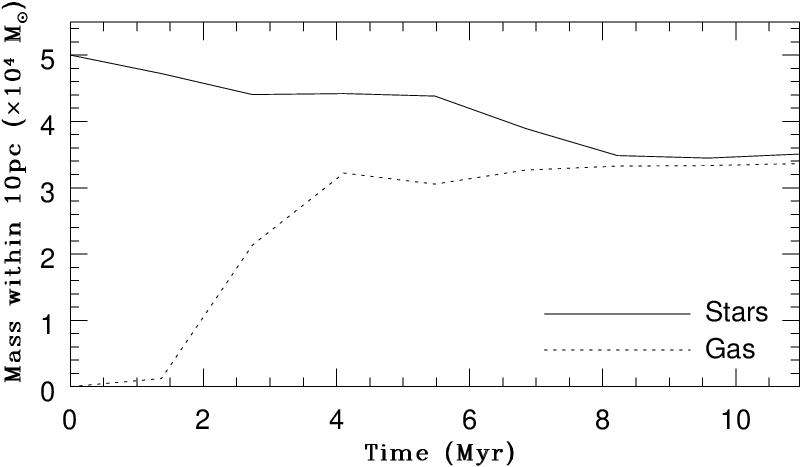,width=8.0cm}
\caption{
The time evolution of stellar (solid) and gaseous (dotted)
masses within the central 10pc of the SC in the standard model M1. 
}
\label{Figure. 2}
\end{figure}

\begin{figure}
\psfig{file=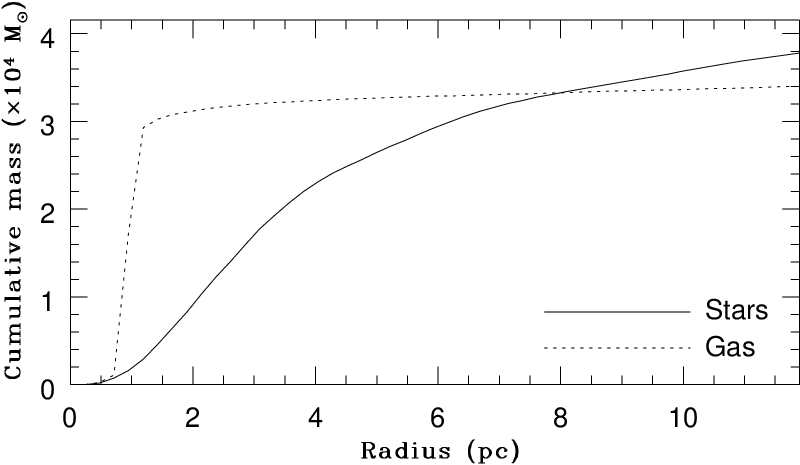,width=8.0cm}
\caption{
The cumulative stellar (solid) and gaseous (dotted) masses 
within radius $R$ from the center of the SC 
at $T=11.0$ Myr in the standard model M1.
}
\label{Figure. 3}
\end{figure}

\begin{figure}
\psfig{file=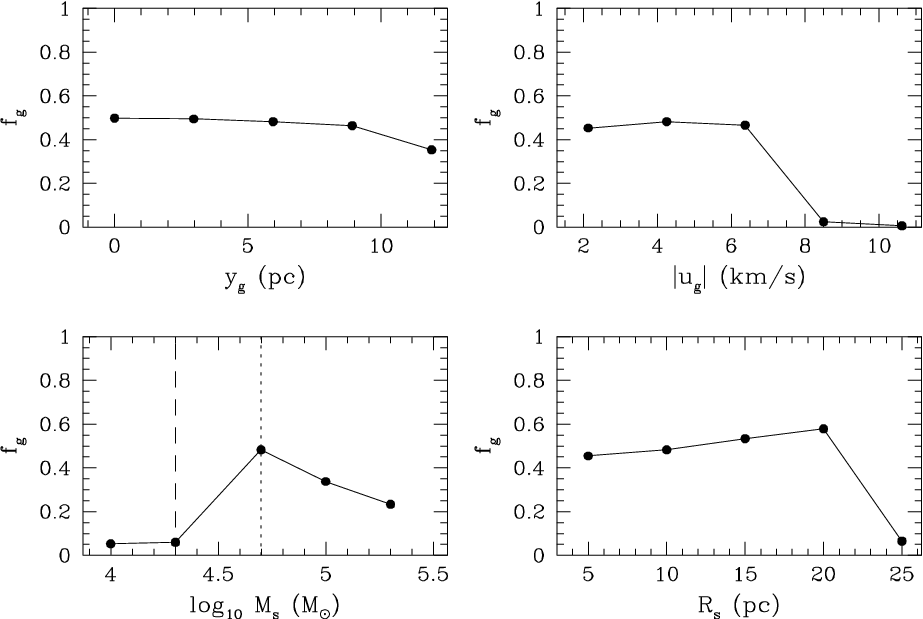,width=8.0cm}
\caption{
Dependences of $f_{\rm g}$ 
(i.e., gas mass fraction within a radius of 10 pc in merger remnants)
on $y_{\rm g}$ (upper left), $|u_{\rm g}|$ (upper right),
$M_{\rm s}$ (lower left), and $R_{\rm s}$ (lower right).
The dotted and dashed lines represent $s_{\rm g}=1$ and $2.5$,
respectively. 
}
\label{Figure. 4}
\end{figure}

\section{Results}

\subsection{The standard model}

Fig. 1 shows how gas in a GMC evolves during major merging
between the GMC and a SC with $s_{\rm g}=1$ in the standard
model M1. During the off-center collision between the GMC and the SC,
the central part of the GMC is tidally compressed to form
a compact gaseous core. 
Owing to dynamical friction during merging,
the core can sink into the central region of the SC to finally form
a flattened gaseous spheroid (or gas disk) within the SC.
The SC is not destroyed by this merging and thus retains
its initial spherical shape,
though it loses about 30\% of its initial mass owing
to tidal stripping.
Hence, the newly formed SC has a very compact, flattened gaseous disk
in its central region: further star formation from this central gas 
can result in two distinct stellar populations in the SC.

About 30\% of the gas in the GMC is rapidly stripped during merging
to form a very diffuse gaseous halo around the new SC. 
This stripped gas cannot be accreted onto the SC 
after merging even in this isolated model,
and thus is likely to be returned back to interstellar medium
(ISM) of the LMC if interaction between the stripped GMC gas and 
the ISM in the LMC is included in future simulations. 
Fig. 2 shows that the gas mass fraction within the central
10 pc of the new SC ($f_{\rm g}$) increases as
the SC loses its stellar mass and gas is accreted into the SC.
Fig. 2 also shows that $f_{\rm g}$
can be finally as large as 0.5 at $T=11$ Myr. 
This implies that if all of the gas can be converted into
new stars, the mass ratio of the second generation of stars
to the first can be as large as 1. We discuss later important
implications of this result in terms of the origin of the DMSTOs.

Fig. 3 shows that the final cumulative radial mass distributions ($M(<R)$, 
where $R$ is the distance from the mass center of the newly-formed 
SC) of gas and stars at $T=11$ Myr are quite different
in the sense that the gas shows a more compact distribution than do the
stars. The half-mass radius is 3.3 pc for the stars and 1.0 pc for the
gas, which reflects that the collapsed central part of the GMC
is directly transfered to the center of the SC owing to
dynamical friction. 
Therefore the gas mass fraction is significantly larger in the inner region
of the new SC, which means that this region of the new
SC would be dominated by young stars if star formation were to 
subsequently occur in the accumulated gas.

\begin{figure}
\psfig{file=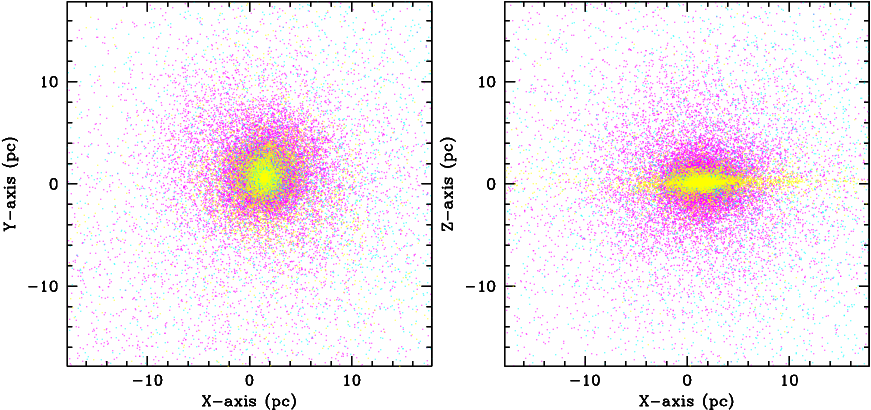,width=8.5cm}
\caption{
Final distributions of stars (magenta), gas (cyan), and new stars
(yellow) projected onto the $x$-$y$ plane (left) and $x$-$z$ plane
(right) for the star formation model M2, in which the model parameters 
are exactly the same as those adopted in
the standard model, but star formation is included.
}
\label{Figure. 5}
\end{figure}

\subsection{Parameter dependences}

The dependences of the final properties of SCs 
on a parameter are  investigated by changing the values of the parameter
and
{\it by fixing those  of other  parameters}.
Dependences of the present numerical results on
the four key parameters ($y_{\rm g}$, $|u_{\rm g}|$, $M_{\rm s}$,
and $R_{\rm s}$) are described as follows:

(i) As shown in Fig. 4, $f_{\rm g}$ does not depend strongly on
$y_{\rm g}$ (i.e., impact parameter)
as long as GMCs merge with SCs (i.e., $y_{\rm g} \le 12$
pc). However $f_{\rm g}$ can be very small ($<0.1$) for models 
with large $y_{\rm g}$, in which no merging occurs and only small fractions 
of gas are transferred to SCs during the GMC-SC tidal interaction.
For example, the model with $y_{\rm g} = 18$ pc 
and other parameter values being the same as those in the standard model
shows $f_{\rm g}=0.07$, corresponding to a gas mass within the central
10 pc of $3.2 \times 10^3 {\rm M}_{\odot}$.

(ii) The models with high $|u_{\rm g}|$ ($ > 7$ km s$^{-1}$)
show very small $f_{\rm g}$ ($<0.1$), because
high-speed encounters between GMCs and SCs can again prevent
GMCs from merging with the SCs (see Fig. 4).
There is no strong dependence of $f_{\rm g}$ on
$|u_{\rm g}|$ for models with $|u_{\rm g}| \le 7$ km s$^{-1}$
in which GMC-SC merging does occur.
Models with  $|u_{\rm g}| >  12 $ km s$^{-1}$
show very small $f_{\rm g}$ ($<0.01$) even if $y_{\rm g}=0$ pc.

(iii) GMCs can either destroy SCs or tidally strip most of their
stellar envelopes, if the masses of the GMCs are significantly larger
than those of the SCs interacting with the GMCs.
Therefore the compact gas cores of GMCs cannot sink into the
central regions of SCs in models with large 
$s_{\rm g}$ ($=M_{\rm g}/M_{\rm s}$).
As a result of this, the models with large $s_{\rm g}$ ($> 2.5$)
show small $f_{\rm g}$ in the merger remnants for a reasonable
set of other model parameters (see Fig. 4).
It should be, however, stressed that the models with $y_{\rm g}=0$,
small $|u_{\rm g}|$ ($<4$ km s$^{-1}$),
and large $s_{\rm g}$ ($> 2.5$) show large $f_{\rm g}$ ($>0.7$),
because in these situations the SCs become trapped in the central
cores of the GMCs. It might not be appropriate to call these very 
gas-dominated merger remnants ``SCs''.

(iv) In models with smaller $s_{\rm g}$ (larger $M_{\rm s}$),
SCs can accumulate a larger amount of gas from merging GMCs.
However, in these models it is not possible for $s_{\rm g}$ to 
be particularly large, owing to the initially larger stellar masses
in these models. Therefore, there is an
optimum $s_{\rm g}$ for which $f_{\rm g}$ is a maximum
for a given set of model parameters.  
As shown in Fig. 4, the major merger model (i.e., $s_{\rm g}=1$)
shows the largest $f_{\rm g}$, which implies that GMC-SC mergers
with comparable masses can form larger fractions of the second
generations of stars after merging.

(v) The models with larger $R_{\rm s}$ show larger $f_{\rm g}$
for $R_{\rm s} \le 20$ pc. This is mainly because
the final stellar masses within the central 10 pc for SCs
with larger $R_{\rm s}$ (and thus lower stellar densities) 
tend to be small owing to the more efficient stripping of stars
during GMC-SC merging: the actual masses of gas within the 
merger remnants do not depend strongly on $R_{\rm s}$
for $R_{\rm s} \le 20$ pc.
The models with large $R_{\rm s}$ (e.g., $R_{\rm s}=25$ pc)
show very small $f_{\rm g}$ owing to the total destruction of
SCs during GMC-SC merging.

\subsection{Star formation}

Owing to our adopted prescription for star formation in GMCs,
new stars are formed preferentially in rather high-density
regions of GMCs. New stars can therefore be formed during
the accumulation of compact and flattened gas spheres 
in the central regions of GMCs due to
strong tidal compression of the GMCs by SCs.
The total mass of new stars within a merger remnant 
(i.e., newly-formed SC) depends on how much gas
is accumulated into the SC during merging.
Therefore the parameter dependences
of mass fractions of new stars in merger remnants
for models including star formation
are very similar to
those of $f_{\rm g}$ for models without star formation.
Thus we here describe only
the results of the standard star formation model M2.

Fig. 5 shows that the new stars in the merger
remnant show a compact, disky distribution
and thus have quite different spatial distributions from the 
original stars: there appears to be a SC within a SC.
This ``SC-within-SC'' appearance is one of key characteristics
of the remnants of SC-GMC mergers in the present study.
The total masses of stars, gas, and new stars
are $3.7 \times 10^4 {\rm M}_{\odot}$,
$1.1 \times 10^4 {\rm M}_{\odot}$,
and $2.0 \times 10^4 {\rm M}_{\odot}$, respectively,
which means that the mass fraction of new stars among all 
stars is 0.35.  
The half-mass radii for stars, gas, and new stars
are 4.0 pc, 3.8 pc, and 1.2 pc, respectively,
which clearly demonstrates that
the new stars have a much more compact distribution than 
do the original stars.
The simulated flattened structure of new  stars
is rather similar to the inner flattened ``core''
observed in M15 (e.g., van den Bosch et al. 2006).

Fig. 6 shows that the star formation rate (SFR) becomes
as high as 0.05 ${\rm M}_{\odot}$ yr$^{-1}$ in the star
formation model M2 when the compact gas sphere
of the GMC is sinking
into the SC owing to dynamical friction.
The SFR is quite peaked so that there is little age
dispersion among the new stars formed during GMC-SC merging.
The model with larger $y_{\rm g}$ (=11.9 pc) and thus a weaker
tidal field from the SC shows a smaller peak SFR 
($\sim 0.02 {\rm M}_{\odot}$ yr$^{-1}$), in the later phase of 
GMC-SC merging.
These results clearly suggest that the strength of the tidal field
of a SC merging with a GMC is a key factor that can determine
the peak SFR and the star formation efficiency of the GMC-SC merger.
Fig. 7 confirms  that the inner regions of the merger remnant
are dominated by new  stars, as suggested in Fig. 5.

The derived nested spatial distributions 
(i.e., SC-within-SC appearance) of the simulated SCs 
are apparently not observed in the intermediate-age SCs of the 
LMC which have been tested (e.g., MBN07).
Furthermore, observations have not yet discovered
inner disky structures of SCs in the LMC,
though the SCs themselves have rather flattened shapes in general
(e.g., van den Bergh 2000).
To resolve these inconsistencies in structural properties 
between the simulated and observed SCs requires that 
long-term dynamical relaxation processes play a role in 
dynamically mixing
old and new components and thus in forming 
the canonical radial density
profiles observed in the present SCs (e.g., King-type profiles).

The median relaxation time in
a SC like NGC 1846 should be $\sim 2$ Gyr (see for example the models
in Mackey et al. 2008b),
and a factor of a few shorter than this in the very centre.
Therefore, there
has conceiveably been enough time for dynamical relaxation
processes to influence and modify
cluster structure. It should also be noted that
the second generation of stars in a composite SC formed as described 
above will initiate a new phase of violent
relaxation in the SC as the most massive members evolve quickly and 
die, and any residual gas is expelled. This can be very good at 
dynamically mixing the cluster on a
relatively short time scale (see e.g., Meylan \& Heggie 1997 and
references therein). It is likely that these evolutionary 
processes might alter significantly the centrally concentrated spatial 
distribution of the second generation of stars, as well as the 
observable number fraction of the first and second generations of 
stars in the centres of new SCs.

\begin{figure}
\psfig{file=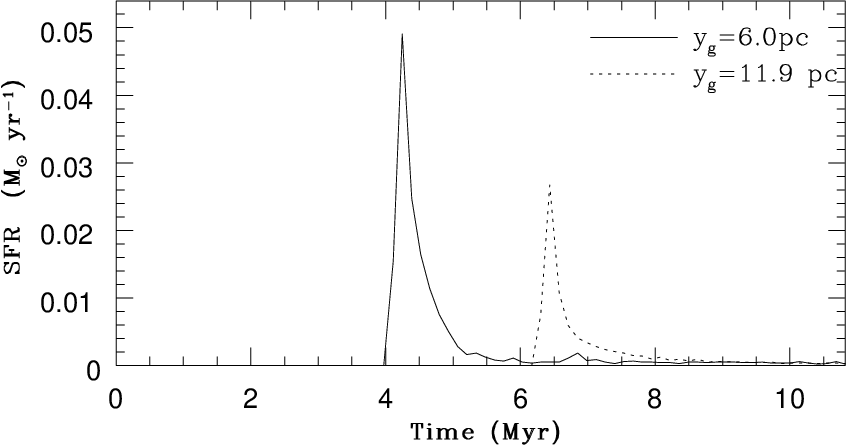,width=8.0cm}
\caption{
The time evolution of star formation rates (SFRs) in units
of ${\rm M}_{\odot}$ yr$^{-1}$ for 
star formation models  with $y_{\rm g}=6.0$ pc
(solid) and $y_{\rm g}=11.9$ pc (dotted). For these models,
gas can be converted into new stars if the gas particles satisfy
the requisite conditions for star formation as described in the 
main text.
Parameter values other than $y_{\rm g}$ for these
two models are exactly the same
as those of the standard model M1. 
}
\label{Figure. 6}
\end{figure}

\begin{figure}
\psfig{file=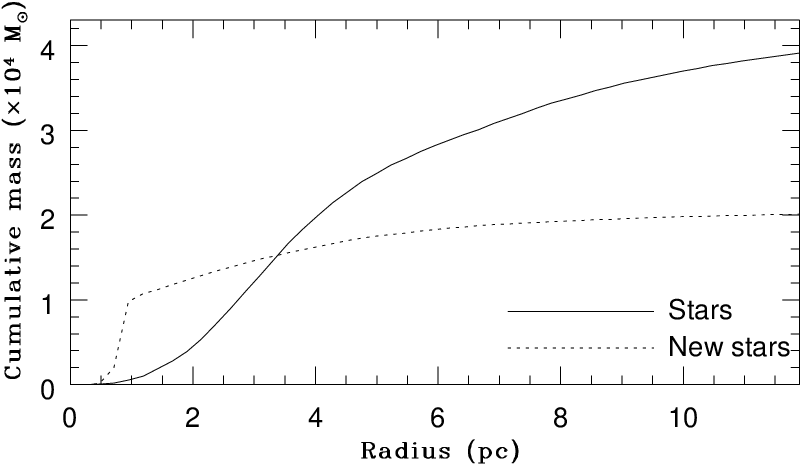,width=8.0cm}
\caption{
The same as Fig. 3 but for stars (solid) and new stars (dotted)
in the standard star formation model M2. 
}
\label{Figure. 7}
\end{figure}

\section{Discussion}

\subsection{Advantages and disadvantages of the scenario}

The observed apparently clear distinction between bluer and redder
MSTOs in some SCs of the LMC (e.g., MBN07; Mackey et al. 2008; 
Milone et al. 2008)
implies that there were two bursts of star formation
rather than prolonged periods of star formation in these SCs.
One advantage of the present scenario is therefore that
it can naturally explain the possible two burst epochs of star formation:
one is during the original formation of SCs and the other is 
during GMC-SC merging.
Furthermore, the new stars formed during the GMC-SC merging 
are clearly bound to the original SC, which naturally explains 
how two different stellar populations may coexist in some SCs.
The scenario does not need to consider a physical mechanism
by which interstellar gas can be accumulated within SCs and then form
stars within the SCs.

The present scenario predicts significant age differences between
the first and the second generations of stars in SCs owing
to the late formation of stars triggered by GMC-SC merging/interaction.
A key question here is therefore whether the age difference can be 
similar to $\sim 3 \times 10^8$ yr 
($t_{\rm gap}$), which may be required to explain the origin of
the DMSTOs (e.g., MBN07; Mackey et al. 2008; 
Milone et al. 2008).
The time scale of a GMC-SC merger event ($t_{\rm m}$) 
in the LMC can be estimated
as follows (e.g., Makino \& Hut 1997):
\begin{equation}
t_{\rm m}=\frac{ 1 } {n_{\rm g}\sigma v},
\end{equation}
where $n_{\rm g}$, $\sigma$, and $v$ 
are the mean number density of the GMCs,
the geometrical cross section of a GMC,
and the relative velocity between a GMC and a SC.
If we consider that the LMC is a uniform disk with 
a disk radius of $r_{\rm d}$ ($ \sim 5$ kpc),
a scale height of $z_{\rm d}$, 
and the number of GMCs being equal to $N_{\rm g}$ for simplicity,
then
\begin{equation}
n_{\rm g} = \frac { N_{\rm g} } {2 \pi z_{\rm f} {r_{\rm d}}^2 } .
\end{equation}
We assume here that SCs are more compact than GMCs with comparable
masses and thus that $\sigma$ in the $t_{\rm m}$ estimation
should be the cross section for GMCs. 
If we consider that GMCs have spherical shapes, then we can use
the following:
\begin{equation}
\sigma = \pi {r_{\rm g}}^2,
\end{equation}
where $r_{\rm g}$ is the size of a GMC (i.e., $r_{\rm g}=0.5 R_{\rm g}$).

We are interested in the SCs of the LMC
a few Gyr ago when the SCs with DMSTOs were formed. The LMC would have 
a larger gas mass fraction and thus a larger number of GMCs in the 
past.
Recent NANTEN observations have found 272 GMCs with masses larger
than $2\times 10^4 M_{\odot}$ in the LMC (Fukui et al. 2008); however 
the number of GMCs a few Gyr ago may be significantly
larger than that in the present LMC.
If we consider (i) that $v$ is similar to the observed velocity
dispersion of stars in the LMC
(e.g., van der Marel 2002), (ii) that the past
$N_{\rm g}$ is larger by a factor of 2 
than that of the present LMC,
and (iii) that $z_{\rm d}$ is about 10\% of the observed radial scale 
length of the LMC (=1.5 kpc; van den Bergh 2000),
then $t_{\rm m}$ can be estimated as follows:
\begin{equation}
t_{\rm m}= 3.2
{ ( \frac{ N_{\rm g} } {500} )  }^{-1}
{ ( \frac{ z_{\rm d} } {150 {\rm pc} } )  }
{ ( \frac{ r_{\rm g} } {15 {\rm pc}} )  }^{-2}
{ ( \frac{ v } {20 {\rm km s^{-1} } } )  }^{-1}
{\rm Gyr} .
\end{equation}
It should be stressed here that the above $t_{\rm m}$
is for the average number density of GMCs
with masses larger than $2 \times 10^4 {\rm M}_{\odot}$ over
the entire LMC region: $t_{\rm m}$ can be much shorter
than the above in some regions with locally high number densities.
The above value of $t_{\rm m}$ is much larger
than $t_{\rm gap}$ ($\sim 0.3$ Gyr) required for explaining
the DMSTOs.
However, $t_{\rm m}$ can be as small as $t_{\rm gap}$
in the LMC if we adopt a higher $N_{\rm g}$, 
an $r_{\rm g}$ larger than equation (1) describes,
and a thinner gas disk as follows:
\begin{equation}
t_{\rm m}= 0.3
{ ( \frac{ N_{\rm g} } {1000} )  }^{-1}
{ ( \frac{ z_{\rm d} } {100 {\rm pc} } )  }
{ ( \frac{ r_{\rm g} } {30 {\rm pc}} )  }^{-2}
{ ( \frac{ v } {20 {\rm km s^{-1} } } )  }^{-1}
{\rm Gyr} .
\end{equation}
It remains unclear which of the above two estimations 
is more reasonable and realistic in the LMC a few Gyr ago.
In addition, owing to differences in the structural and kinematical 
properties of gas and stars between the LMC and the SMC 
(i.e., $N_{\rm g}$, $z_{\rm d}$, $r_{\rm g}$, $r_{\rm d}$, and  $v$),
$t_{\rm m}$ might well be different between the MCs.

A SC can merge with a GMC located within a distance $d_{\rm g}$ from
the SC within the following time scale:
\begin{equation}
t_{\rm m}= 0.3
{ ( \frac{ d_{\rm g} } {66 {\rm pc}} )  }^{2}
{ ( \frac{ z_{\rm d} } {150 {\rm pc} } )  }
{ ( \frac{ r_{\rm g} } {15 {\rm pc}} )  }^{-2}
{ ( \frac{ r_{\rm g} } {20 {\rm km s^{-1} } } )  }^{-1}
{\rm Gyr} .
\end{equation}
This means that $t_{\rm m}$ can be similar
to $t_{\rm gap}$ for some local regions with {\rm surface} number
densities (${\Sigma}_{\rm g}$) of GMCs similar to 73 kpc$^{-2}$. 
Such high ${\Sigma}_{\rm g}$ would be possible in strong
tidal arms and in the inner region of the LMC.
It is also possible that a group of GMCs (or GMC associations) 
could have such a high ${\Sigma}_{\rm g}$ in the LMC.
Thus, $t_{\rm m}$ can be as small as $\sim 3\times 10^8$ yr 
(thus $\sim t_{\rm gap}$), although it is possible that $t_{\rm m}$ 
varies significantly between different regions of the LMC a few Gyr ago.

It appears
that the observed possible age differences between two
populations is an order of $10^8$ yr: no SCs with DMSTOs
appear to show implied age differences of an order of $10^9$ yr
(see Santiago et al. 2002 for a possibly  exceptional case of NGC 1868).
Therefore one possible problem in the present scenario is that
the time scales of GMC-SC merging
can {\it in principle} be
significantly longer than $10^8$ yr for some SCs:
the scenario needs to explain why there are apparently no
intermediate-age SCs with implied age differences between the two 
component stellar populations being as large as 1 Gyr. 
A possible solution of this potential problem is if the
time scales of GMC-SC merging in the LMC grew longer
in the more recent past -- that is, as the LMC evolved 
with time GMC-SC merging became less likely 
because the number density of GMCs and and the typical GMC size 
became lower and smaller, respectively, owing to
rapid gas consumption by ongoing star formation in the LMC.
Thus it would be possible that 
intermediate-age SCs cannot merge with GMCs 
much later ($\sim 1$ Gyr) than their formation in the LMC.

Furthermore, it would be equally possible that the time scales of 
GMC-SC merging can be significantly shorter than $10^8$ yr owing
to locally high number densities of GMCs in the vicinity of some SCs.
Although SCs which experience merging with GMCs much less than $10^8$ yr
after their formation can still have two distinct component populations,
they would not clearly show DMSTOs in their CMDs due
to the much smaller age differences between them.
Therefore the possible presence of SCs with age differences 
significantly smaller than $10^8$ yr
would be still consistent with observations. 
Merging of SCs with GMCs less than $10^8$ yr after their formation
might well be responsible for the observed binary/multiple SCs with small
age differences (e.g., NGC 1850, Leon et al. 1999).

GMC-SC merging is highly likely to occur between SCs and their
local neighborhood GMCs in the LMC: cloud complexes
(or ``superclouds'') with the typical
mass of $10^7 {\rm M}_{\odot}$ (Elmegreen 1987) would be
the progenitors for the GMC-SC mergers.  
Therefore differences in chemical abundances between merging SCs
and GMCs should be quite small unless there exist significant radial and 
azimuthal abundance gradients in the LMC.
Owing to the time lag of $\sim 3 \times 10^8$ yr between SC formation
and SC-GMC merging in the GMC-SC merger scenario,
GMCs can be chemically polluted by AGB stars in SCs 
merging with the GMCs.
Therefore, it is quite possible that the second generations of stars 
formed from GMCs could have different abundances in light
elements from those in the first generations initially in the SCs.
Ongoing spectroscopic observations on the abundance properties
of SCs with DMSTOs will soon reveal whether or not 
abundance inhomogeneities exist in these objects.
We thus plan to investigate the chemical evolution of GMC-SC mergers
in order to compare between the predicted and the observed  
degrees of inhomogeneity in light element abundances 
for SCs with DMSTOs in the LMC.

\subsection{Alternative scenarios}
\subsubsection{Self-pollution by AGB stars}

It would be possible that stellar ejecta from AGB stars in a SC
can be accumulated in the central
region of the SC so that the ejecta can be used for the formation
of the second generation of stars.  
In this ``self-pollution'' scenario, 
the second generation of stars needs to form well after the
removal of stellar ejecta from Type II supernovae that can cause
a significant spread in heavy element abundances.
Recent numerical simulations of GC formation based on this self-pollution
scenario have shown (i) that star formation for the second generation
can start soon after ejection of stellar winds from massive AGB stars,
and (ii) that star formation can continue {\it gradually} 
within GCs over a period of $\sim 100$ Myr (D'Ercole et al. 2008).

Thus the self-pollution
scenario has serious difficulties in explaining (i) why the time 
difference between the formation epochs
of the first and second generations of stars is typically $\sim 300$ Myr, 
and (ii) why most stars in the second generation formed almost 
simultaneously.
It is likely that the numerical simulations by D'Ercole et al. (2008)
do not describe so precisely the star formation histories within
SCs owing to the adopted rather idealized models for hydrodynamics and star
formation within SCs.
Therefore, future theoretical studies
based on more sophisticated numerical simulations are
certainly worthwhile to confirm whether the self-pollution 
scenario has really the above serious problems.

\subsubsection{Star cluster merging}

It is a  well known observational fact that a significant fraction
of SCs in the LMC are binary or multiple clusters  
(e.g., Bhatia \& Hatzidimitriou 1988;
Bhatia et al. 1991; Dieball et al. 2002).
This fact implies that merging of binary SCs can form single
SCs with distinct stellar populations, if there are initial differences 
in age between the two original clusters. 
This merger scenario needs to explain why age differences
between two merging clusters can be similar to $3 \times 10^8$ yr
for some binary SCs.
Observational studies on possible age differences in binary
SCs in the LMC have demonstrated 
that the components of binary or multiple SC systems typically 
appear to be small, hence implying that all SCs in a given
bound system are generally formed simultaneously or over a very short 
time scale (e.g., Dieball et al. 2002).
Therefore, the merger scenario appears to have a serious problem
in explaining the frequently occurring age difference of $\sim 3\times 10^8$ 
yr observed for the DMSTOs in intermediate-age MC clusters.

It should, however, be stressed that there are some {\it binary} 
SCs in the LMC with age differences 
up to $\sim 0.5$ Gyr (e.g., SL356-357 pair,  Leon et al. 1999):
it would be possible that merging of the two component SCs 
in such systems can happen in the LMC.
If this is the case, the merger scenario needs to explain
(i) why SC merging apparently happens {\it preferentially}
for those systems which have age differences
of $\sim 300$ Myr and (ii) why the mass-ratios
of the younger SCs to the older ones   
are comparable to and larger than 1.
It is currently unclear whether such preferential SC merging
can happen in the history of the LMC.

\subsection{DMSTOs only for intermediate-age SCs?}

The age distribution of the LMC
SCs shows a gap extending from 3 to 13 Gyr --
with only one cluster in this age range --
suggesting that a second epoch of cluster formation started abruptly 
in the LMC
about 3 Gyr ago
(e.g., Da Costa 1991;  Geisler et al. 1997; Rich et al. 2001;  Piatti et
al. 2002).
SCs with DMSTOs are observed to be $\sim 1.5-2.5$ Gyr old 
(Mackey et al. 2008; Milone et al. 2008) and thus were 
formed just after the ``age-gap'' period.  
So far it remains observationally unclear whether the oldest GCs in the
LMC, which formed before the commencement of the age gap, 
also show DMSTOs.
This raises the following two questions: (i) whether the origin of
the SCs with DMSTOs is closely associated with some specific formation
processes at the reactivation of SC formation a few Gyr
ago, and (ii) whether SCs can have DMSTOs irrespective of 
their formation epochs owing to a general physical mechanism
responsible for the DMSTO formation.

Recent numerical simulations have shown that tidal interaction between
the LMC and the SMC can dramatically increase (by a factor of ten)
the cloud-cloud collision rate in the LMC during the interaction
(e.g., see Fig. 1 in Bekki et al. 2004a).
This result implies that if  
a high cloud-cloud collision rate means a high GMC-SC
collision/merging rate in the interacting MC system,
the origin of SCs with DMSTOs might be closely associated
with the commencement of strong tidal interaction between the MCs
a few Gyr ago in the present GMC-SC merger scenario.
Recent numerical simulations have suggested that 
the MCs might have started their strong tidal interaction
about $3-4$ Gyr ago for a reasonable set of
orbital parameters (Bekki et al. 2004b; Bekki \& Chiba 2005).
Thus it is possible that the origin of
SCs with DMSTOs may result from high GMC-SC collision/merging
rates in the LMC  due to strong tidal interactions
between the MCs a few Gyr ago.

Owing to continuous interaction between the MCs after
the last dynamical coupling about $3-4$ Gyr ago 
(e.g., Bekki \& Chiba 2005), the LMC can retain an enhanced 
cloud-cloud (and thus probably GMC-SC) collision rate.
This means that SCs with ages younger than 1.5 Gyr can also
show DMSTOs, though strong observational evidence for the
presence of DMSTOs in the CMDs of young SCs does not yet exist.
Previous numerical simulations (e.g., Yoshizawa \& Noguchi 2003)
have shown that the LMC and SMC can very strongly interact with 
each other about 1.5 Gyr ago -- in this case it would be highly 
likely that 
a larger fraction of SCs with ages of 1.5$\pm0.3$ Gyr
may show DMSTOs. This would be consistent with the recent
observational results of Milone et al. (2008).
In addition, the cloud-SC collision rate at the epoch
of disk formation in the LMC might well be much higher 
than that at the present owing to the higher gas mass fraction
and the higher degree of random motion at the time of disk formation.
Therefore, it is very possible that the old GCs in the LMC may
show DMSTOs in the GMC-SC merger scenario. 

It should be stressed, however, that
recent proper motion measurements of the MCs by the
Advanced Camera for Surveys (ACS) on the  {\it Hubble
Space Telescope (HST)} have reported that
the LMC and the SMC have significantly high Galactic
tangential velocities ($367 \pm 18$ km s$^{-1}$
and $301 \pm 52$ km s$^{-1}$, respectively), suggesting that 
the MCs may be unbound
from each other (e.g., Kallivayalil et al. 2006).
In this case, LMC-SMC interactions about 1.5 
and $3-4$ Gyr ago are unlikely to have occurred, meaning that the 
formation of SCs with DMSTOs
has little to do with any past interaction between the MCs.
Given that detailed modeling of the LMC-SMC orbits 
that includes the new proper motion data as well as factors previously 
ignored 
(e.g., a common halo and/or different circular velocity of the Galaxy)
has not yet been completed,
it would be fair to say that at present the connection between 
the interaction histories of the MCs, and the formation of DMSTOs 
in SCs is not so clear.

\subsection{Relations to the Galactic GCs with abundance
inhomogeneities?}

The origin of the observed star-to-star light element abundance 
inhomogeneities in the Galactic GCs (GGCs)
has been extensively discussed
both theoretically and observationally (e.g., Sneden et al. 1992;
Norris \& Da Costa 1995; Smith et al. 2005;
see Gratton et al. 2004 for a recent review).
The observed presence of star-to-star abundance inhomogeneities
in less evolved stars
on the main sequence and subgiant-branch (e.g., Cannon et al. 1998
for 47 Tuc) now strongly suggests that the origin of these abundance
inhomogeneities
is due to the early chemical evolution of GC-forming gas clouds.
Stellar ejecta from AGB stars and massive stars has been considered
to play a key role in the early chemical evolution of GCs for
the self-pollution scenario (e.g., Karakas et al. 2006;
Prantzos \& Charbonnel 2006).
The early chemical pollution in GC-forming clouds results from 
stellar ejecta from massive AGB stars 
in the ``AGB scenario''  
(e.g., Ventura \& D'Antona 2005; Bekki et al. 2007)
and from the stellar winds from massive stars 
in the ``massive star scenario'' (e.g., Prantzos \& Charbonnel 2006).

If the origin of the GGCs with abundance inhomogeneities
is essentially the same as that of the SCs with DMSTOs in the LMC,
then age differences between the first and the second generations of
stars in these GGCs may be as large as $\sim 3 \times 10^8$ yr.
This means that the massive star scenario can be ruled out,
because the time lag between the first and the second generations
is at most an order of $\sim 10^6$ yr (Bekki \& Chiba 2007;
Decressin et al. 2007).
This also means that gaseous ejecta from the first generation of 
AGB stars in a SC
needs to be retained in the SC {\it without star formation
for $\sim 3\times 10^8$ yr} and then converted into the second 
generation in the AGB scenario. 
Such delayed formation of the second generation of stars
implies that the second generation would form from the 
mixed gas of stellar ejecta from AGB stars with different masses.
Thus, if only stellar ejecta from massive AGB stars 
($\sim 6 {\rm M}_{\odot}$) can much better
explain the observed C-N, O-Na, and Mg-Al anticorrelations
(e.g., Ventura \& D'Antona et al. 2008),
the AGB scenario would have a serious problem in explaining
the possible age different of $\sim 3 \times 10^8$ yr between the 
first and the second generations.

It would be equally possible that the origin of GCs with abundance 
inhomogeneities in the Galaxy is different from that of SCs
with DMSTOs in the LMC,
given that the physical properties are different between
these clusters in the two different galaxies (e.g., Mackey \& Gilmore 2004).
The implied high fraction of He-rich stars with $Y>0.3$
in $\omega$ Cen and NGC 2808 (Piotto et al. 2005; Piotto et al. 2007) 
cannot be explained simply by the GMC-SC merger scenario: 
such He-rich stars would need to form from stellar ejecta either 
from fast-rotating massive stars
(e.g., Prantzos \& Charbonnel 2006; Decressin et al. 2007)
or from massive AGB stars (e.g., D'Antona et al. 2002).
Possibly, the time-lags between formation of the first and the 
second generations of stars in clusters can be different 
between host galaxies.
This might well cause a variety of differences 
in physical properties between
the first and second generations of stars in clusters 
belonging to separate host galaxies.

\section{Conclusions}

We have numerically investigated 
how GMCs evolve during GMC-SC merging in order
to better understand the origin of the DMSTOs observed in 
intermediate-age SCs in the MCs. We have mainly investigated
mass fractions of gas within 10 pc of the remnants
of GMC-SC mergers for variously different model
parameters. 
We summarise our principle results as
follows:

(1) Gas initially in GMCs can be accumulated within 
SCs during GMC-SC merging so that high-density compact 
gaseous regions that are significantly flattened 
are formed in the central regions of the SCs.
The mass fractions of gas ($f_{\rm g}$) 
within the central 10 pc of the merger remnants 
(i.e., the newly formed SCs)
can be as large as 0.5 for models with $s_{\rm g} \sim 1$
(i.e., major mergers).

(2) There is an optimum $M_{\rm s}$ (or $s_{\rm g}$)
for a given set of model parameters, for which $f_{\rm g}$ 
is a maximum. For example,
the model with $M_{\rm s}=M_{\rm g}$ (i.e., $s_{\rm g}=1$)
shows the maximum $f_{\rm g}$
in models with $M_{\rm g}=5 \times 10^4 {\rm M}_{\odot}$
and $y_{\rm g}=0.5 R_{\rm g}$ (=6 pc).
The remnants of major GMC-SC mergers are likely to show
larger $f_{\rm g}$ in the present study.

(3) As long as GMC-SC merging occurs,
$f_{\rm g}$ does not depend strongly on
$y_{\rm g}$ and $|u_{\rm g}|$. The models with larger
$y_{\rm g}$ and $|u_{\rm g}|$,
for which GMCs can interact with SCs without merging,
show only very small $f_{\rm g}$ (an order of $10^{-2}$). 
These results mean that merging is essential for 
the formation of new SCs with large $f_{\rm g}$.

(4) If star formation is included in GMC-SC merging,
compact, flattened star clusters composed of new stars
can be formed in the central regions of the merger remnants. 
Since the simulated SCs-within-SCs appearances 
(i.e., doubly nested SCs) are not observed in the MCs,
some later dynamical evolution processes need to transform
the doubly nested SCs into normal-looking (well-mixed) objects 
with standard (King-type) density profiles.  

(5) The time scale of GMC-SC merging ($t_{\rm m}$) in the LMC  
can be similar to the typical age difference of $\sim 3 \times 10^8$ 
yr between the component stellar populations implied by observations
of DMSTOs in SCs in the LMC.
However $t_{\rm m}$ depends strongly on the number of GMCs,
the sizes of GMCs, and the velocity dispersions of GMCs and stars 
in the LMC a few Gyr ago, all of which remain observationally unclear.

Based on these results, we have pointed out that the observed possibly 
large fractions of the second generations of stars in SCs with 
DMSTOs in the LMC can be due to the past GMC-SC merging.
We have also suggested that
time lags between SC formation and the subsequent GMC-SC merging
can be responsible for the 
possible age differences between the first and the second
generations of stars in the SCs with DMSTOs in the LMC.
The adopted numerical code does not allow us
to investigate whether the simulated doubly-nested SCs can evolve  
into normal SCs with canonical radial density profiles due to 
internal dynamical processes. 
We plan to investigate this question in our future studies using 
the appropriate numerical codes (e.g., {\tt NBODY4}).

\section{Acknowledgment}
We are  grateful to the anonymous referee for valuable comments,
which contribute to improve the present paper.
KB acknowledges the financial support of the Australian Research
Council
throughout the course of this work.
ADM is supported on a Marie Curie Excellence Grant
from the European Commission under contract MCEXT-CT-2005-025869.


\begin{thebibliography}{}

\bibitem[]{}
Baume, G., Carraro, G., Costa, E., Mendez, R. A., Girardi, L.
2007, MNRAS, 375, 1077

\bibitem[]{}
Bekki, K., Forbes, D. A., Beasley, M. A., Couch, W. J. 
2002, MNRAS, 335, 1176

\bibitem[]{}
Bekki, K., Couch, W. J. 2003, ApJ, 596, L13

\bibitem[]{}
Bekki, K., Beasley, M. A.,  Forbes, D. A.,  Couch, W. J.
2004a, ApJ,  602, 730

\bibitem[]{}
Bekki, K., Couch, W. J., Beasley, M. A.,  Forbes, D. A., Chiba, M., 
Da Costa, G.
2004b, 610, L93 

\bibitem[]{}
Bekki, K.,  Chiba, M. 2005, MNRAS, 356, 680 

\bibitem[]{}
Bekki, K.,  Chiba, M. 2007, ApJ, 665, 1164 

\bibitem[]{}
Bekki, K., Campbell, S. W., Lattanzio, J. C.,  Norris, J. E.
2007, MNRAS, 377, 335

\bibitem[]{}
Bertelli, G., Nasi, E., Girardi, L., Chiosi, C., Zoccali, M.,
Gallart, C. 2003, AJ, 125, 770

\bibitem[]{}
Bhatia, R. K.,  Hatzidimitriou, D. 1988, MNRAS, 230, 215

\bibitem[]{}
Bhatia, R. K., Read, M. A., Tritton, S.,  Hatzidimitriou, D. 1991,
A\&AS, 87, 335

\bibitem[]{}
Bonnor, W. B. MNRAS, 116, 351

\bibitem[]{}
Cannon, R. D.,  Croke, B. F. W.,
Bell, R. A., Hesser, J. E., Stathakis, R. A., 1998, MNRAS, 298, 601

\bibitem[]{}
Charbonnel, C.,  Prantzos, N., 2006, preprint (astro-ph/0606220)


\bibitem[]{}
Da Costa G. S. 1991, 
in Haynes R., Milne D., eds, Proc. IAU Symp. 148,
The Magellanic Clouds, Kluwer, Dordrecht, 
p183

\bibitem[]{}
D'Antona, F., Caloi, V., Montalba\'an, J., Ventura, P., Gratton, R.
2002, A\&A, 395, 69

\bibitem[]{}
Decressin, T., Meynet, G., Charbonnel, C.,
Prantzos, N., Ekstr\"om, S. 2007, A\&A, 464, 1029


\bibitem[]{}
D'Ercole, A., Vesperini, E., D'Antona, F.,  McMillan,
S. L. W., Recchi, S. 2008, preprint (astro-ph/0809.1438)

\bibitem[]{}
Dieball, A.,  Grebel, E. K. 1998, A\&A, 339, 773

\bibitem[]{}
Ebert, R. 1995, ZA, 37, 217

\bibitem[]{}
Elmegreen, B. G. 1987, ApJ, 312, 626

\bibitem[]{}
Fukui, Y. et al. 2008, ApJS, 178, 56

\bibitem[]{}
Geisler, D.,  Bica, E., Dottori, H., Claria, J. J.,
Piatti, A. E.,   Santos, J. F. C., Jr. 1997,
AJ, 114, 1920 


\bibitem[]{}
Glatt, K. et al. 2008, AJ, 136, 1703

\bibitem[]{}
Goudfrooij, P., Puzia, T. H., Kozhurina-Platais, V., Chandar, R. 
2008, AJ, submitted

\bibitem[]{}
Gouliermis, D. A., Lianou, S., Kontizas, M., Kontizas, E., Dapergolas, A. 
2006, ApJ, 652, L93

\bibitem[]{}
Gratton, R.,  Sneden, C.,  Carretta, E., 2004, ARA\&A, 42, 385

\bibitem[]{}
Hurley, J. R.,  Bekki, K. 2008, MNRAS, 389, L61 (HB08)

\bibitem[]{}
Kallivayalil, N., van der Marel, R. P., Alcock, C.
2006, ApJ, 652, 1213 


\bibitem[]{}
Karakas, A. I., Fenner, Y., Sills, A.,
Campbell, S. W.; Lattanzio,
J. C. 2006, ApJ, 652, 1240

\bibitem[]{}
Larson, R. B. 1981, MNRAS, 194, 809

\bibitem[]{}
Leon, S., Bergond, G., Vallenari, A. 1999, A\&A, 344, 450

\bibitem[]{}
Mackey, A. D.,  Broby Nielsen, P. 2007, MNRAS, 379, 151

\bibitem[]{}
Mackey, A. D., Broby Nielsen, P., Ferguson, A. M. N.,
Richardson, J. C. 2008a, ApJ, 681, L17

\bibitem[]{}
Mackey, A. D., Gilmore, G. F. 2004, MNRAS, 355, 504

\bibitem[]{}
Mackey, A. D., Wilkinson, M. I., Davies, M. B.,  Gilmore, G. F.
2008b, MNRAS, 386, 65

\bibitem[]{}
Makino, J., Hut, P. 1997, ApJ, 481, 83

\bibitem[]{}
Meylan, G., Heggie, D. C. 1997, A\&AR, 8, 1

\bibitem[]{}
Milone, A. P.,  Bedin, L. R., Piotto, G., Anderson, J.
2008, preprint (astro-ph/0810.2558)

\bibitem[]{}
Mucciarelli, A., Carretta, E., Origlia, L., Ferraro, F. R. 2008, 
AJ, 136, 375

\bibitem[]{}
Norris, J. E.,  Da Costa, G. S., 1995, ApJ, 441, L81


\bibitem[]{}
Piatti A.,  Sarajedini, A.,  Geisler, D., Bica, E.   Claria, J. J. 
2002, MNRAS, 329, 556



\bibitem[]{}
Piotto, G. 2008, MmSAI, 79, 334

\bibitem[]{}
Piotto, G. et al. 2005, ApJ, 621, 777

\bibitem[]{}
Piotto, G. et al. 2007, ApJ, 661, L53

\bibitem[]{}
Prantzos, N., Charbonnel, C. 2006, A\&A, 458, 135




\bibitem[]{}
Rich, R. Michael.,  Shara, M. M.   Zurek, D.  
2001, AJ, 122, 842

\bibitem[]{}
Santiago, B., Kerber, L., Castro, R., de Grijs, R. 2002, 
MNRAS, 336, 139

\bibitem[]{}
Smith, G. H., Briley, M. M.,  Harbeck, D.,
2005, AJ, 129, 1589



\bibitem[]{}
Sneden, C., Kraft, R. P.,  Prosser, C. F., Langer, G. E.,
1992, AJ, 104, 2121


\bibitem[]{}
Solomon, P. M., Sanders, D. B., Scoville, N. Z.
1979, 
in Proc. IAU Symp. 84, The Large-scale Characteristics of the Galaxy,
ed. W. B. Burton (Dordrecht: Reidel), 35

\bibitem[]{}
van den Bosch, R.  de Zeeuw, T. Gebhardt, K.,  Noyola, E.,  van de
Ven, G. 2006, ApJ, 641, 852

\bibitem[]{}
van den Bergh, S. 2000,
The Galaxies of the Local Group, Cambridge: Cambridge Univ. Press.

\bibitem[]{}
van der Marel, R. P., Alves, D. R., Hardy, E.,  Suntzeff,
N. B. 2002, AJ, 124, 2639

\bibitem[]{}
Ventura, P., D'Antona, F. 2008, A\&A, 479, 805

\bibitem[]{}
Yoshizawa, A.,  Noguchi, M. 2003, MNRAS, 339, 1135 


\end{thebibliography}
\end{document}